\documentclass[10pt,twocolumn,twoside]{IEEEtran}

\usepackage{graphicx}
\usepackage{amsmath}
\usepackage{amssymb}
\usepackage{subcaption}
\usepackage{tikz}
\usepackage{pgfplots}
\pgfplotsset{compat=1.3}
\usetikzlibrary{arrows,decorations,shapes}
\usepackage{balance}

\DeclareMathAlphabet{\mathbit}{OML}{cmr}{bx}{it}

\DeclareMathOperator{\e}{e}
\DeclareMathOperator{\E}{E}
\DeclareMathOperator{\T}{T}

\renewcommand\vec[1]{\operatorname{vec}\left(#1\right)}
\renewcommand\arcsin[1]{\operatorname{arcsin}\left(#1\right)}

\DeclareMathOperator{\fieldR}{\mathbb{R}}

\DeclareMathOperator{\fieldN}{\mathbb{N}}

\newcommand{\ve}[1]{\boldsymbol{#1}}

\newcommand{\exdi}[2]{\E_{#1} \left[#2\right]}

\renewcommand{\exp}[1]{\operatorname{exp}\left(#1\right)}

\newcommand\Sign{\operatorname{sign}}
\newcommand{\signum}[1]{\Sign{\left(#1\right)}}

\newcommand{\sinc}[1]{\operatorname{sinc} \left(#1\right)}

\title{Spectral Power Parameter Estimation of\\Random Sources with Binary Sampled Signals}
\author{Manuel~S.~Stein
\thanks{This work was funded by the Deutsche Forschungsgemeinschaft (DFG, German Research Foundation) - grant no. 413008418.}
\thanks{M. S. Stein is with the Department of Microelectronics, Technische Universiteit Delft, The Netherlands (e-mail: M.S.Stein@tudelft.nl).}
}

\begin{document}
\maketitle
\begin{abstract}
This paper investigates the problem of estimating the spectral power parameters of random analog sources using numerical measurements acquired with minimum digitization complexity. Therefore, spectral analysis has to be performed with binary samples of the analog sensor output. Under the assumption that the structure of the spectral power density of the analog sources is given, we investigate the achievable accuracy for power level estimation with likelihood-oriented processing. The discussion addresses the intractability of the likelihood with multivariate hard-limited samples by exploiting advances on probabilistic modeling and statistical processing of hard-limited multivariate Gaussian data. In addition to estimation-theoretic performance analysis, the results are verified by running an iterative likelihood-oriented algorithm with synthetic binary data for two exemplary sensing setups.
\end{abstract}
\section{Introduction}
Analyzing hard-limited sensor data is a traditional topic in signal processing. Early approaches \cite{Vleck66,Hinich67} were motivated by the lack of computing power, which hindered the processing with high amplitude resolution \cite{Weinreb63, Steinberg75}. In recent work, amplitude resolution minimization has again received attention due to data transmission \cite{Ribeiro06} or analog-to-digital (A/D) conversion constraints \cite{Mezghani07,Mo15,Mollen17,Jacobsson17,Rao19}. Combining both aspects indicates that signal digitization via binary sampling is also key for high-performance sensing systems. The simplicity of basic digital processing steps, small sensor data volume, and energy-efficient analog front-end design make it possible to deploy higher temporal, spectral, and spatial sampling rates. Such, with the same A/D resources, digital measurements can be acquired, which contain more information about the analog signal model \cite{Stein19}. To extract all this information from the data, efficient binary signal processing methods are required.

In this context, likelihood-oriented spectral analysis with noisy binary measurements is discussed. We assume that the analog signal is a superposition of ideally band-limited Gaussian processes, each characterized by a spectral power density of unknown weight. In practice, such an assumption is never exactly fulfilled. However, in several applications, signals approximately follow such additive random models \cite{Schulz97}. Besides, within the class of distributions with the same covariance structure, the Gaussian one has maximum entropy \cite{CoverBook}. This theoretical argument supports the practical relevance of the Gaussian assumption by suggesting that it forms a conservative modeling perspective. Here the analog signal is digitized via a low-complexity binary sampling device while the digital sensor data shall be processed optimally. This requires characterization of the likelihood which is intractable for multivariate binary distributions. We address this by using a binary distribution model with reduced sufficient statistics which provides access to a conservative version of the Fisher information matrix \cite{Stein_FisherBound}. Based on this framework, we investigate the sensitivity gap in comparison to sampling the analog signal with high A/D resolution and verify the results by Monte-Carlo simulations of an iterative estimation algorithm.

Note that, direct estimation of the spectrum function from hard-limited data is considered in \cite{Vleck66,Hinich67,Weinreb63} using an asymptotic relationship between quantized and unquantized autocorrelation. Approximate maximum-likelihood estimators (MLEs) for the parameters of an autoregressive process are derived in \cite{Kedem80} by exploiting the asymptotic independence between distant samples of a Gaussian process. In contrast, here the estimation of spectral parameters characterizing the analog random process is considered under an auxiliary likelihood applicable also to samples of small size. For the distinct problem of estimating the parameters of deterministic sinusoidal signals in white noise from hard-limited samples, see e.g., \cite{Madsen00}.

\section{System Model}
For the discussion, we assume an analog sensor signal
\begin{align}\label{analog:sensor:signal}
y(t)=\sum_{d=1}^{D} x_{d}(t)+\eta(t), \quad y(t)\in\fieldR,t\in\fieldR, 
\end{align}
comprising $D\in\fieldN$ source signals
\begin{align}
x_{d}(t)=x_{\text{c},d}(t)\cos{(t\omega_d)}+x_{\text{s},d}(t)\sin{(t\omega_d)},
\end{align}
each with an individual frequency $\omega_d\in\fieldR$. The components $x_{c/s,d}(t)\in\fieldR$ are random processes with power spectral density $\Psi_d(\omega)=\theta_d, \theta_d>0,$ for $\omega\in [ -\Omega_d;  \Omega_d]$ and $\Psi_d(\omega)=0$ elsewhere. The auto-correlation of such random processes is
\begin{align}
r_d(t)&=\exdi{x_{\text{c}/\text{s}}}{x_{\text{c}/\text{s},d}(\tau) x_{\text{c}/\text{s},d}(\tau-t)}\notag\\
&= \frac{1}{2\pi} \int_{-\Omega_d}^{\Omega_d} \Psi_d(\omega) \e^{{\rm j} \omega t} {\rm d}\omega\notag\\
&=\theta_d\frac{\Omega_d}{\pi} \sinc{\frac{\Omega_d}{\pi} t}.
\end{align}
In \eqref{analog:sensor:signal}, $\eta(t)\in\fieldR$ denotes independent additive noise which is also modeled as an ideally band-limited random process with $\Psi_0(\omega)=\theta_0$ for $\omega\in [ -\Omega_0;  \Omega_0]$ and $\Psi_0(\omega)=0$ elsewhere. 

\subsection{Digital data model - Ideal sampling}
Sampling \eqref{analog:sensor:signal} at a rate of $f_{\text{A/D}}=\frac{\Omega_{\text{A/D}}}{\pi}$ for a duration of $T_{\text{A/D}}=\frac{M}{f_{\text{A/D}}}, M\in\fieldN,$ with an A/D converter featuring $\infty$-bit amplitude resolution, results in $M$-variate samples of the form
\begin{align}\label{digital:samples}
\ve{y}=\sum_{d=1}^{D}  \ve{x}_{d}+ \ve{\eta}, \quad \ve{y},\ve{x}_{d},\ve{\eta}\in\fieldR^{M},
\end{align}
with a covariance matrix of structure
\begin{align}
\ve{R}_{\ve{y}}(\ve{\theta})&=\exdi{\ve{y};\ve{\theta}}{\ve{y}\ve{y}^{\T}}\notag\\
&=\sum_{d=1}^{D} \theta_d \frac{\Omega_d}{\pi} \ve{\Sigma}_{d}\odot\ve{W}_{d}+\theta_0 \frac{\Omega_0}{\pi}\ve{\Sigma}_{0},\\
\label{parameter:vector}
\ve{\theta}&=\begin{bmatrix} \theta_1 &\ldots &\theta_D &\theta_0\end{bmatrix}^{\T}. 
\end{align}
The entries of the source and noise correlation matrices are
\begin{align}
\left[ \ve{\Sigma}_{d} \right]_{ij}= \sinc{\frac{\Omega_d}{\Omega_{\text{A/D}}} |i-j|},\quad i,j=1,\ldots, M,
\end{align}
while the entries of the mixing matrices are
\begin{align}
[ \ve{W}_{d} ]_{ij}&=\cos{\bigg(\frac{\omega_{d}}{\Omega_{\text{A/D}}}\pi(i-1) \bigg)}\cos{\bigg(\frac{\omega_{d}}{\Omega_{\text{A/D}}}\pi(j-1) \bigg)}\notag\\ &+\sin{\bigg(\frac{\omega_{d}}{\Omega_{\text{A/D}}}\pi(i-1) \bigg)}\sin{\bigg(\frac{\omega_{d}}{\Omega_{\text{A/D}}}\pi(j-1) \bigg)}.
\end{align}
Additionally, we define the correlation matrix
\begin{align}\label{correlation:matrix}
\ve{\Sigma}_{\ve{y}}(\ve{\theta})
&=\frac{\sum_{d=1}^{D} \theta_d \frac{\Omega_d}{\pi} \ve{\Sigma}_{d}\odot\ve{W}_{d}+\theta_0 \frac{\Omega_0}{\pi}\ve{\Sigma}_{0}}{\sum_{d=1}^{D} \theta_d \frac{\Omega_d}{\pi} +\theta_0 \frac{\Omega_0}{\pi}}.
\end{align}
Samples \eqref{digital:samples} are assumed to follow the Gaussian distribution
\begin{align}\label{distribution:gaussian}
p_{\ve{y}}(\ve{y};\ve{\theta})=\frac{\exp{-\frac{1}{2} \ve{y}^{\T} \ve{R}_{\ve{y}}^{-1}(\ve{\theta}) \ve{y}}}{ \sqrt{(2\pi)^{M} \det{(\ve{R}_{\ve{y}}(\ve{\theta}))}} }.
\end{align}

\subsection{Digital data model - Binary sampling}
If the signal \eqref{analog:sensor:signal} is digitized via a low-complexity A/D converter with $1$-bit amplitude resolution, the measurements are
\begin{align}\label{hard:limiter}
\ve{z}=\signum{\ve{y}},
\end{align}
where the element-wise hard-limiter $\signum{\cdot}$ is defined
\begin{align}
\left[\ve{z}\right]_m=
\begin{cases}
+1& \text{if } [\ve{y}]_m \geq 0,\\
-1 & \text{if } [\ve{y}]_m < 0.
\end{cases}
\end{align}
The output of \eqref{hard:limiter} is invariant to changes of its input scale, such that one can fix $\theta_0=1$ and reduce \eqref{parameter:vector} by one element. 
Like \eqref{distribution:gaussian}, binary distributions are exponential families
\begin{align}\label{distribution:exponential:family}
{p}_{\ve{z}}(\ve{z};\ve{\theta})=\exp{\ve{w}^{\T}(\ve{\theta}) \ve{\phi}(\ve{z}) - \lambda(\ve{\theta})+\nu(\ve{z})},
\end{align}
where $\ve{w}(\ve{\theta})\colon \fieldR^{D} \to \fieldR^{C}$ are the statistical weights, $\ve{\phi}(\ve{z})\colon \ve{\mathcal{Z}} \to\fieldR^{C}$ the sufficient statistics, $\lambda(\ve{\theta})\colon \fieldR^D \to \fieldR$ the log-normalizer, and $\nu(\ve{z})\colon \ve{\mathcal{Z}} \to\fieldR$ the carrier measure. In contrast to \eqref{distribution:gaussian}, where $C\in\mathcal{O}(M^2)$, in multivariate binary distributions $C\in\mathcal{O}(2^M)$ as, beside the pairwise products, the sufficient statistics also comprise all higher order products \cite{Dai13}. Therefore, we use an auxiliary exponential family model \cite{Stein_FisherBound}
\begin{align}\label{aux:distribution:exponential:family}
\tilde{p}_{\ve{z}}(\ve{z};\ve{\theta})=\exp{\ve{\tilde{w}}^{\T}(\ve{\theta}) \ve{\tilde{\phi}}(\ve{z}) - \tilde{\lambda}(\ve{\theta})+\tilde{\nu}(\ve{z})},
\end{align}
where the sufficient statistics $\ve{\tilde{\phi}}(\ve{z})\colon \ve{\mathcal{Z}} \to\fieldR^{\tilde{C}}$ are
\begin{align}\label{aux:statistics:quantizer}
\ve{\tilde{\phi}}(\ve{z})&=\ve{\Phi}\vec{\ve{z} \ve{z}^{\T}}
\end{align}
with $\ve{\Phi}\in [0; 1]^{\tilde{C} \times M^2}$ being an elimination matrix canceling the duplicate and constant elements of $\ve{z} \ve{z}^{\T}$. This reduces \eqref{distribution:exponential:family} to a quadratic distribution \cite{Cox94} for which $\tilde{C}\in\mathcal{O}(M^2)$. If
\begin{align}
\exdi{\ve{{z}};\ve{\theta}}{\ve{ \tilde{\phi}}(\ve{z}) }&=\exdi{\ve{\tilde{z}};\ve{\theta}}{\ve{ \tilde{\phi}}(\ve{z}) },\\
\exdi{\ve{{z}};\ve{\theta}}{ \ve{ \tilde{\phi}}(\ve{z}) \ve{\tilde{\phi}}^{\T}(\ve{z}) }&=\exdi{\ve{\tilde{z}};\ve{\theta}}{\ve{ \tilde{\phi}}(\ve{z}) \ve{ \tilde{\phi} }^{\T}(\ve{z})},
\end{align}
where $\exdi{\ve{z};\ve{\theta}}{\cdot}$ denotes the expectation under \eqref{distribution:exponential:family} and $\exdi{\ve{\tilde{z}};\ve{\theta}}{\cdot}$ under \eqref{aux:distribution:exponential:family}, the Fisher matrices of \eqref{distribution:exponential:family} and \eqref{aux:distribution:exponential:family} satisfy
\begin{align}
\ve{F}_{\ve{z}}(\ve{\theta}) \succeq \ve{\tilde{F}}_{\ve{z}}(\ve{\theta}).
\end{align}
Without explicit characterization of \eqref{distribution:exponential:family}, the information contained in samples from \eqref{distribution:exponential:family} is conservatively quantified by
\begin{align}
\ve{\tilde{F}}_{\ve{z}}(\ve{\theta})=\bigg(\frac{\partial \ve{\mu}_{\ve{\tilde{\phi}}}(\ve{\theta})}{ \partial \ve{\theta}} \bigg)^{\T} \ve{R}_{\ve{\tilde{\phi}}}^{-1}(\ve{\theta}) \frac{\partial \ve{\mu}_{\ve{\tilde{\phi}}}(\ve{\theta})}{ \partial \ve{\theta}}
\end{align}
where
\begin{align}\label{mean:aux:statistics}
\ve{\mu}_{\ve{\tilde{\phi}}}(\ve{\theta})&=\exdi{\ve{z};\ve{\theta}}{\ve{ \tilde{\phi}}(\ve{z}) }=\ve{\Phi}\vec{\ve{R}_{\ve{z}}(\ve{\theta})}
\end{align}
is obtained by the arcsine law \cite{ThomasBook}
\begin{align}\label{arcsine:law}
\ve{R}_{\ve{z}}(\ve{\theta})&=\frac{2}{\pi} \arcsin{\ve{\Sigma}_{\ve{y}}(\ve{\theta})}.
\end{align}
Therefore, the $d$th column of the derivative of \eqref{mean:aux:statistics} is
\begin{align}
\left[ \frac{\partial \ve{\mu}_{\ve{\tilde{\phi}}}(\ve{\theta})}{\partial \ve{\theta}} \right]_{d} =\ve{\Phi} \vec{ \frac{\partial \ve{R}_{\ve{z}}(\ve{\theta}) }{\partial \theta_d}}
\end{align}
with the off-diagonal matrix entries
\begin{align}\label{quantizer:covariance:derivative:entries}
&\left[\frac{\partial \ve{R}_{\ve{z}}(\ve{\theta}) }{\partial \theta_{d} }\right]_{ij}=\frac{2}{\pi} \frac{ \left[\frac{\partial \ve{\Sigma}_{\ve{y}}(\ve{\theta}) }{\partial \theta_{d} }\right]_{ij} }{ \sqrt{ 1 - \left[ \ve{\Sigma}_{\ve{y}}(\ve{\theta})  \right]^2_{ij}} }, \quad \forall i,j: i\neq j,
\end{align}
while the diagonal entries are zero. With matrix \eqref{correlation:matrix},
\begin{align}
\frac{\partial \ve{\Sigma}_{\ve{y}}(\ve{\theta}) }{\partial \theta_{d} }&=\frac{ \frac{\Omega_d}{\pi} \big(\ve{\Sigma}_{d}\odot\ve{W}_{d} - \ve{\Sigma}_{\ve{y}}(\ve{\theta}) \big)}{\sum_{d'=1}^{D} \theta_{d'} \frac{\Omega_{d'}}{\pi} + \frac{\Omega_0}{\pi}}.
\end{align}
The covariance matrix of the auxiliary statistics \eqref{aux:statistics:quantizer} 
\begin{align}\label{covariance:aux:statistics}
\ve{R}_{\ve{\tilde{\phi}}}(\ve{\theta})&=\exdi{\ve{{z}};\ve{\theta}}{ \ve{ \tilde{\phi}}(\ve{z}) \ve{\tilde{\phi}}^{\T}(\ve{z}) } - \ve{\mu}_{\ve{\tilde{\phi}}}(\ve{\theta})\ve{\mu}^{\T}_{\ve{\tilde{\phi}}}(\ve{\theta})
\end{align}
is obtained by evaluating expectations $\exdi{\ve{z};\ve{\theta}}{z_i z_j z_k z_l}$ which, besides \eqref{arcsine:law}, requires quadrivariate orthant probabilities \cite{Sinn11}.

\section{Processing Task}
Given $N\in\fieldN$ independent samples of the binary output \eqref{hard:limiter}
\begin{align}\label{one:bit:data}
\ve{Z}=\begin{bmatrix} \ve{z}_1 &\ve{z}_2 &\ldots &\ve{z}_N\end{bmatrix},
\end{align}
the optimum technique is the MLE. As the binary likelihood \eqref{distribution:exponential:family} is intractable, parameter estimation is here performed by
\begin{align}\label{definition:cmle}
\ve{\hat{\theta}}(\ve{Z}) &= \arg \max_{\ve{\theta}\in\ve{\Theta}} \sum_{n=1}^{N} \ln \tilde{p}_{\ve{z}}(\ve{z}_n;\ve{\theta}).
\end{align}
The solution of \eqref{definition:cmle} consistently achieves \cite{Stein_FisherBound}
\begin{align}\label{definition:error:mle:quant}
\ve{R}_{\ve{\hat{\theta}}}(\ve{\theta})&=\exdi{\ve{Z};\ve{\theta}}{ \big(\ve{\hat{\theta}}(\ve{Z})-\ve{\theta}\big) \big(\ve{\hat{\theta}}(\ve{Z})-\ve{\theta}\big)^{\rm{T}} }\notag\\
&\overset{a}{=}\frac{1}{N} \ve{\tilde{F}}_{\ve{z}}^{-1}(\ve{\theta})
\end{align}
and, after computing the empirical mean statistics
\begin{align}
\ve{\hat{\mu}_{\ve{\tilde{\phi}}}}(\ve{Z})=\frac{1}{N}\sum_{n=1}^{N} \ve{\tilde{\phi}}(\ve{z}_n),
\end{align}
can be found quickly by $I\in\fieldN$ iterations of scoring \cite{Mak93}
\begin{align}\label{scoring:rule:quant}
\ve{\hat{\theta}}^{(i)}&=\ve{\hat{\theta}}^{(i-1)}+\Delta \ve{\hat{\theta}}(\ve{Z};\ve{\hat{\theta}}^{(i-1)})
\end{align}
with the update term \cite{Stein_MARLENE}
\begin{align}\label{scoring:update:quant}
\Delta \ve{\hat{\theta}}(\ve{Z};\ve{\theta})
&= \Bigg( \bigg(\frac{\partial \ve{\mu}_{\ve{\tilde{\phi}}}(\ve{\theta})}{ \partial \ve{\theta}} \bigg)^{\T} \ve{R}_{\ve{\tilde{\phi}}}^{-1}(\ve{\theta}) \frac{\partial \ve{\mu}_{\ve{\tilde{\phi}}}(\ve{\theta})}{ \partial \ve{\theta}} \Bigg)^{-1}\notag\\
&\phantom{=}\cdot \bigg( \frac{\partial \ve{\mu}_{\ve{\tilde{\phi}}}(\ve{\theta})}{ \partial \ve{\theta}} \bigg)^{\T} \ve{R}_{\ve{\tilde{\phi}}}^{-1}(\ve{\theta})\big(\ve{\hat{\mu}}_{\ve{\tilde{\phi}}}(\ve{Z}) - \ve{\mu}_{\ve{\tilde{\phi}}}(\ve{\theta})\big).
\end{align}
After each update \eqref{scoring:rule:quant}, element-wise back-projection $\ve{\hat{\theta}}^{(i)}=\operatorname{max}(\theta_\Delta,\ve{\hat{\theta}}^{(i)})$ with $\theta_\Delta>0$ ensures that $\ve{\hat{\theta}}^{(i)}\in \ve{\Theta}$.
\section{Results}
For performance analysis, we define the relative measure
\begin{align}\label{hardlimiting:loss}
\chi_d=\frac{ \left [ \ve{F}_{\ve{y}}^{-1}(\ve{\theta}) \right ]_{dd}}{ \left [ \ve{\tilde{F}}_{\ve{z}}^{-1}(\ve{\theta})\right ]_{dd}},
\end{align}
where $\ve{F}_{\ve{y}}(\ve{\theta})$ denotes the Fisher information matrix of the unquantized samples \eqref{digital:samples}. For verification by Monte-Carlo simulations with the iterative approach \eqref{scoring:rule:quant}, we define
\begin{align}\label{uncertainty}
\hat{\sigma}_d &= \frac{ 1 }{ \theta_d } \left [ \ve{\hat{R}}^{\frac{1}{2}}_{\ve{\hat{\theta}}}(\ve{\theta}) \right ]_{dd}
\end{align}
with the empirical error covariance matrix
\begin{align}
\ve{\hat{R}}_{\ve{\hat{\theta}}}(\ve{\theta})&= \frac{1}{K} \sum_{k=1}^{K} \big(\ve{\hat{\theta}}(\ve{Z}_k)-\ve{\theta}\big) \big(\ve{\hat{\theta}}(\ve{Z}_k)-\ve{\theta}\big)^{\rm{T}},
\end{align}
where $K$ denotes the number of realizations of the data \eqref{one:bit:data}.

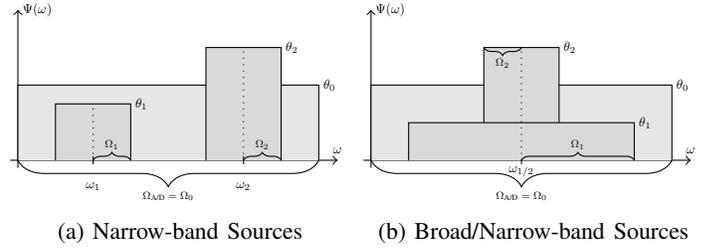
\begin{figure}[!ht]
\centering
    \begin{subfigure}[t]{0.25\textwidth}
    
    \begin{tikzpicture}[scale=0.5, every node/.style={transform shape}]

 	\draw[->] (-0.2,0) -- (8.5,0) node[above,yshift=2pt] {$\omega$};
  	\draw[->] (0,-0.2) -- (0,4) node[right] {$\Psi(\omega)$};
  
  	\draw[fill=gray!20] (0,0) --(0,2) -- (8,2) -- (8,0);
  	\draw [decorate,decoration={brace,amplitude=10pt,mirror}] (0.0,0) -- (8,0) node [black,midway,below,yshift=-20pt] {\footnotesize $\Omega_{\text{A/D}}=\Omega_0$};
  	\draw (8,2) node[right] {$\theta_0$};
    
  	\draw[shift={(4-2,0)}, fill=gray!30] (-1,0) -- (-1,1.5) -- (1,1.5) -- (1,0);
   	\draw[shift={(4-2,0)}, style=dotted] (0,0) -- (0,1.5);
  	\draw[shift={(4-2,0)}] (0pt,3pt) -- (0pt,-3pt) node[below,yshift=-10pt] {$\omega_1$};
   	\draw [decorate,decoration={brace,amplitude=3pt},shift={(4-2,0)}] (0,0) -- (1,0) node [black,midway,above,yshift=5pt] {\footnotesize $\Omega_1$};
   	\draw[shift={(4-2,0)}, fill=gray!30] (1,1.5) node[right] {$\theta_1$};
  
 	\draw[shift={(4+2,0)}, fill=gray!30] (-1,0) -- (-1,3) -- (1,3) -- (1,0);
  	\draw[shift={(4+2,0)}, style=dotted] (0,0) -- (0,3);
  	\draw[shift={(4+2,0)}] (0pt,3pt) -- (0pt,-3pt) node[below,yshift=-10pt] {$\omega_2$};
  	\draw [decorate,decoration={brace,amplitude=3pt},shift={(4+2,0)}] (0,0) -- (1,0) node [black,midway,above,yshift=5pt] {\footnotesize $\Omega_2$};
  	\draw[shift={(4+2,0)}, fill=gray!30] (1,3) node[right] {$\theta_2$};

\end{tikzpicture}
\caption{Narrow-band Sources}
\label{figure:examples:two:narrow}
\end{subfigure}%
    ~ 
\begin{subfigure}[t]{0.25\textwidth}

\begin{tikzpicture}[scale=0.5, every node/.style={transform shape}]

 	\draw[->] (-0.2,0) -- (8.5,0) node[above,yshift=2pt] {$\omega$};
  	\draw[->] (0,-0.2) -- (0,4) node[right] {$\Psi(\omega)$};
  
  	\draw[fill=gray!20] (0,0) --(0,2) -- (8,2) -- (8,0);
  	\draw [decorate,decoration={brace,amplitude=10pt,mirror}] (0.0,0) -- (8,0) node [black,midway,below,yshift=-20pt] {\footnotesize $\Omega_{\text{A/D}}=\Omega_0$};
  	\draw (8,2) node[right] {$\theta_0$};
    
 	\draw[shift={(4,0)}, fill=gray!30] (-1,0) -- (-1,3) -- (1,3) -- (1,0);
  	\draw[shift={(4,0)}, style=dotted] (0,0) -- (0,3);
  	\draw[shift={(4,0)}] (0pt,3pt) -- (0pt,-3pt) node[below,yshift=2pt] {$\omega_{1/2}$};
  	\draw [decorate,decoration={brace,amplitude=3pt},shift={(3,3)}] (1,0) -- (0,0) node [black,midway,below,yshift=-5pt] {\footnotesize $\Omega_2$};
  	\draw[shift={(4,0)}, fill=gray!30] (1,3) node[right] {$\theta_2$};
	
  	\draw[shift={(4,0)}, fill=gray!30] (-3,0) -- (-3,1) -- (3,1) -- (3,0);
   	\draw[shift={(4,0)}, style=dotted] (0,0) -- (0,1);
   	\draw [decorate,decoration={brace,amplitude=3pt},shift={(4,0)}] (0,0) -- (3,0) node [black,midway,above,yshift=5pt] {\footnotesize $\Omega_1$};
   	\draw[shift={(4,0)}, fill=gray!30] (3,1) node[right] {$\theta_1$};
  
\end{tikzpicture}
\caption{Broad/Narrow-band Sources}
\label{figure:examples:broad:narrow}
\end{subfigure}%
\caption{Example Scenarios}
\label{figure:examples}
\end{figure}

Two exemplary setups, with $D=2$, $M=64$, $N=10^5$, and $\Omega_{\text{A/D}}=\Omega_{0}$ are considered. In the first scenario, qualitatively depicted in Fig. \ref{figure:examples:two:narrow}, sources feature a narrow relative bandwidth $\bar{\Omega}_{d}=\frac{\Omega_{d}}{\Omega_{0}}=\frac{1}{64}$ while $\bar{\omega}_{1}=\frac{\omega_{1}}{\Omega_{0}}=\frac{1}{4}$ and $\bar{\omega}_{2}=\frac{3}{4}$. Such a situation can occur, for example, in spectrum monitoring of mobile radio frequencies \cite{Yucek09}. 
Fig. \ref{two:narrow:loss:vs:snr:diff} visualizes the information loss due to hard-limiting \eqref{hardlimiting:loss} as a function of  $\bar{\theta}_{2}=\frac{{\theta}_{2}}{{\theta}_{0}}$ while $\bar{\theta}_{1}$ is fixed. It can be observed that the loss $\chi_1$ decreases when switching from $\bar{\theta}_{1}=-15$ dB to $\bar{\theta}_{1}=-3$ dB. The loss $\chi_2$ stays the same in both situations and obtains its minimum around $\bar{\theta}_{2}=12.6$ dB. The plot shows that a large signal-to-noise ratio (SNR) imbalance is unfavorable for the weaker source while the loss can be below $\chi_d=-2$ dB.
\pgfplotsset{legend style={rounded corners=2pt,nodes=right}}
\begin{figure}[!ht]
\begin{center}
\begin{tikzpicture}[scale=1.0]

  	\begin{axis}[ylabel=$\text{Information Loss }\chi \text{ [dB]}$,
  			xlabel=$\text{Signal-to-Noise Ratio } \bar{\theta}_{2}$,
			grid,
			ymin=-9,
			ymax=-1,
			xmin=-15,
			xmax=20,
			legend pos= south west,
			height = 190pt,
			width = 240pt]
			
			\addplot[blue, style=dashed, line width=0.75pt, every mark/.append style={solid, fill=blue},mark repeat=2, mark=square*] table[x index=0, y index=1]{Data/TwoNarrow_vs_SNR_SNR1_-15_QLOSSdB.txt};
			\addlegendentry{$\chi_1 (\bar{\theta}_{1}=-15 \text{ dB})$};
			
			\addplot[red, style=dashed, line width=0.75pt, every mark/.append style={solid, fill=red},mark repeat=2, mark=triangle*] table[x index=0, y index=2]{Data/TwoNarrow_vs_SNR_SNR1_-15_QLOSSdB.txt};
			\addlegendentry{$\chi_2 (\bar{\theta}_{1}=-15 \text{ dB})$};
			
			\addplot[blue, style=dotted, line width=0.75pt, every mark/.append style={solid, fill=blue},mark repeat=2, mark phase=2, mark=*] table[x index=0, y index=1]{Data/TwoNarrow_vs_SNR_SNR1_-3_QLOSSdB.txt};
			\addlegendentry{$\chi_1 (\bar{\theta}_{1}=\,\,\,-3 \text{ dB})$};
			
			\addplot[red, style=dotted, line width=0.75pt, every mark/.append style={solid, fill=red},mark repeat=2, mark phase=2, mark=diamond*] table[x index=0, y index=2]{Data/TwoNarrow_vs_SNR_SNR1_-3_QLOSSdB.txt};
			\addlegendentry{$\chi_2 (\bar{\theta}_{1}=\,\,\,-3 \text{ dB})$};
																	
	\end{axis}

\end{tikzpicture}
\caption{Information Loss (Narrow Sources)}
\label{two:narrow:loss:vs:snr:diff}
\end{center}
\end{figure}
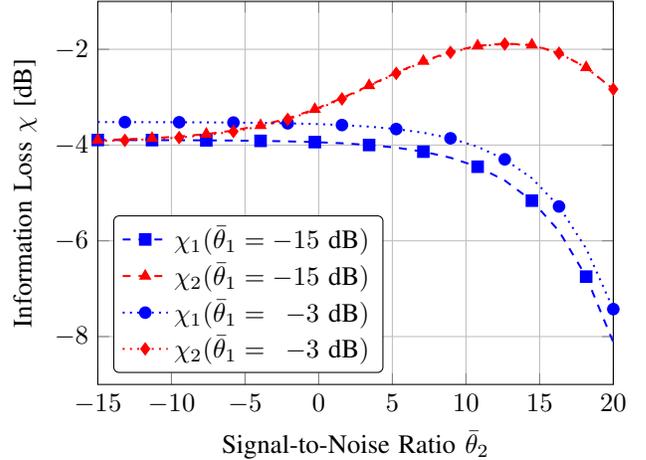
Fig. \ref{two:narrow:nrmse:vs:snr} shows \eqref{uncertainty} under $\bar{\theta}_{1}=-12$ dB when scoring \eqref{scoring:rule:quant} with $I=5$, $\hat{\theta}^{(0)}_d=\theta_\Delta=-30$ dB, and $K=10^3$. The empirical results match the analytical results obtained from the inverse Fisher information matrices (dashed and dotted lines). 
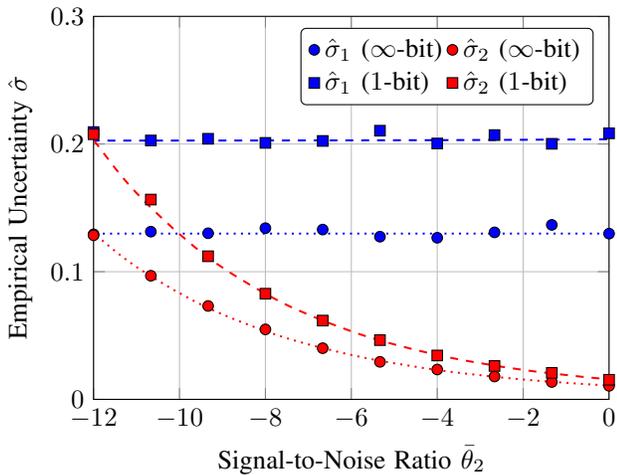
\begin{figure}[!ht]
\begin{center}
\begin{tikzpicture}[scale=1.0]

  	\begin{axis}[ylabel=$\text{Empirical Uncertainty } \hat{\sigma}$,
  			xlabel=$\text{Signal-to-Noise Ratio } \bar{\theta}_{2}$,
			grid,
			ymin=0,
			ymax=0.3,
			xmin=-12,
			xmax=0,
			legend columns=2, 
			legend pos= north east,
			height = 190pt,
			width = 240pt]
			
			\addplot[only marks, every mark/.append style={solid, fill=blue}, mark=*] table[x index=0, y index=1]{Data/TwoNarrow_vs_SNR_Sim_Ideal_SNR1_-12_REL_1000_nRMSE.txt};
			\addlegendentry{$\hat{\sigma}_1$ ($\infty$-bit)};
		        
		         \addplot[only marks, every mark/.append style={solid, fill=red}, mark=*] table[x index=0, y index=2]{Data/TwoNarrow_vs_SNR_Sim_Ideal_SNR1_-12_REL_1000_nRMSE.txt};
		         \addlegendentry{$\hat{\sigma}_2$ ($\infty$-bit)};
		         
		         \addplot[only marks, every mark/.append style={solid, fill=blue}, mark=square*] table[x index=0, y index=1]{Data/TwoNarrow_vs_SNR_Sim_Quant_SNR1_-12_REL_1000_nRMSE.txt};
		         \addlegendentry{$\hat{\sigma}_1$ ($1$-bit)};
		         
		         \addplot[only marks, every mark/.append style={solid, fill=red}, mark=square*] table[x index=0, y index=2]{Data/TwoNarrow_vs_SNR_Sim_Quant_SNR1_-12_REL_1000_nRMSE.txt};
		         \addlegendentry{$\hat{\sigma}_2$ ($1$-bit)};
			
			\addplot[blue, style=dotted, line width=0.75pt, smooth] table[x index=0, y index=1]{Data/TwoNarrow_vs_SNR_SNR1_-12_nRMSE.txt};
	
			\addplot[blue, style=dashed, line width=0.75pt, smooth] table[x index=0, y index=2]{Data/TwoNarrow_vs_SNR_SNR1_-12_nRMSE.txt};

			\addplot[red, style=dotted, line width=0.75pt, smooth] table[x index=0, y index=3]{Data/TwoNarrow_vs_SNR_SNR1_-12_nRMSE.txt};

			\addplot[red, style=dashed, line width=0.75pt, smooth] table[x index=0, y index=4]{Data/TwoNarrow_vs_SNR_SNR1_-12_nRMSE.txt};
															
	\end{axis}

\end{tikzpicture}
\caption{Uncertainty (Narrow Sources; $\bar{\theta}_{1}=-12 \text{ dB}$)}
\label{two:narrow:nrmse:vs:snr}
\end{center}
\end{figure}

\begin{figure}[!ht]
\begin{center}
\begin{tikzpicture}[scale=1.0]

  	\begin{axis}[ylabel=$\text{Information Loss }\chi \text{ [dB]}$,
  			xlabel=$\text{Signal-to-Noise Ratio } \bar{\theta}_{2}$,
			grid,
			ymin=-9,
			ymax=-1,
			xmin=-15,
			xmax=20,
			legend pos= south west,
			height = 190pt,
			width = 240pt]
			
			\addplot[blue, style=dashed, line width=0.75pt, every mark/.append style={solid, fill=blue},mark repeat=2, mark=square*] table[x index=0, y index=1]{Data/BroadNarrow_vs_SNR_SNR1_-15_QLOSSdB.txt};
			\addlegendentry{$\chi_1 (\bar{\theta}_{1}=-15 \text{dB})$};
			
			\addplot[red, style=dashed, line width=0.75pt, every mark/.append style={solid, fill=red},mark repeat=2, mark=triangle*] table[x index=0, y index=2]{Data/BroadNarrow_vs_SNR_SNR1_-15_QLOSSdB.txt};
			\addlegendentry{$\chi_2 (\bar{\theta}_{1}=-15 \text{dB})$};
			
			\addplot[blue, style=dotted, line width=0.75pt, every mark/.append style={solid, fill=blue},mark repeat=2, mark phase=2, mark=*] table[x index=0, y index=1]{Data/BroadNarrow_vs_SNR_SNR1_-3_QLOSSdB.txt};
			\addlegendentry{$\chi_1 (\bar{\theta}_{1}=\,\,-3 \text{dB})$};
			
			\addplot[red, style=dotted, line width=0.75pt, every mark/.append style={solid, fill=red},mark repeat=2, mark phase=2, mark=diamond*] table[x index=0, y index=2]{Data/BroadNarrow_vs_SNR_SNR1_-3_QLOSSdB.txt};
			\addlegendentry{$\chi_2 (\bar{\theta}_{1}=\,\,-3 \text{dB})$};
																
	\end{axis}

\end{tikzpicture}
\caption{Information Loss (Broad/Narrow Sources)}
\label{broad:narrow:loss:vs:snr:diff}
\end{center}
\end{figure}
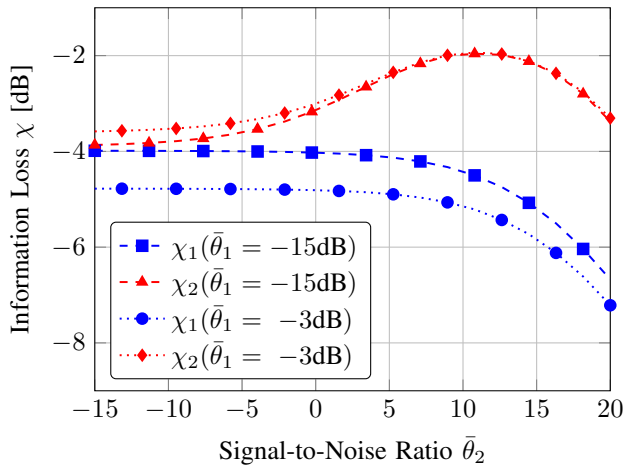

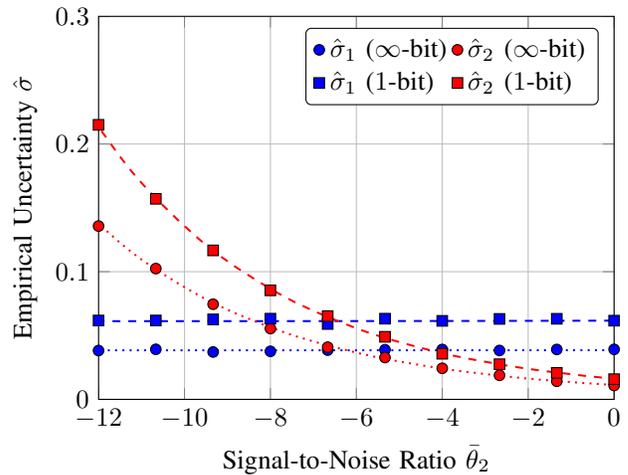
\begin{figure}[!ht]
\begin{center}
\begin{tikzpicture}[scale=1.0]

  	\begin{axis}[ylabel=$\text{Empirical Uncertainty } \hat{\sigma}$,
  			xlabel=$\text{Signal-to-Noise Ratio } \bar{\theta}_{2}$,
			grid,
			ymin=0,
			ymax=0.3,
			xmin=-12,
			xmax=0,
			legend columns=2, 
			legend pos= north east,
			height = 190pt,
			width = 240pt]
			
			\addplot[only marks, every mark/.append style={solid, fill=blue}, mark=*] table[x index=0, y index=1]{Data/BroadNarrow_vs_SNR_Sim_Ideal_SNR1_-12_REL_1000_nRMSE.txt};
			\addlegendentry{$\hat{\sigma}_1$ ($\infty$-bit)};
		        
		         \addplot[only marks, every mark/.append style={solid, fill=red}, mark=*] table[x index=0, y index=2]{Data/BroadNarrow_vs_SNR_Sim_Ideal_SNR1_-12_REL_1000_nRMSE.txt};
		         \addlegendentry{$\hat{\sigma}_2$ ($\infty$-bit)};
		         
		         \addplot[only marks, every mark/.append style={solid, fill=blue}, mark=square*] table[x index=0, y index=1]{Data/BroadNarrow_vs_SNR_Sim_Quant_SNR1_-12_REL_1000_nRMSE.txt};
		         \addlegendentry{$\hat{\sigma}_1$ ($1$-bit)};
		         
		         \addplot[only marks, every mark/.append style={solid, fill=red}, mark=square*] table[x index=0, y index=2]{Data/BroadNarrow_vs_SNR_Sim_Quant_SNR1_-12_REL_1000_nRMSE.txt};
		         \addlegendentry{$\hat{\sigma}_2$ ($1$-bit)};
		         
			\addplot[blue, style=dotted, line width=0.75pt, smooth] table[x index=0, y index=1]{Data/BroadNarrow_vs_SNR_SNR1_-12_nRMSE.txt};
			
			\addplot[blue, style=dashed, line width=0.75pt, smooth] table[x index=0, y index=2]{Data/BroadNarrow_vs_SNR_SNR1_-12_nRMSE.txt};

			\addplot[red, style=dotted, line width=0.75pt, smooth] table[x index=0, y index=3]{Data/BroadNarrow_vs_SNR_SNR1_-12_nRMSE.txt};
			
			\addplot[red, style=dashed, line width=0.75pt, smooth] table[x index=0, y index=4]{Data/BroadNarrow_vs_SNR_SNR1_-12_nRMSE.txt};				
	\end{axis}

\end{tikzpicture}
\caption{Uncertainty (Broad/Narrow Sources; $\bar{\theta}_{1}=-12 \text{ dB}$)}
\label{broad:narrow:nrmse:vs:snr}
\end{center}
\end{figure}

As a second example, we consider a situation with one broad-band signal source featuring relative bandwidth $\bar{\Omega}_{1}=\frac{1}{4}$ and one narrow-band interfering source with $\bar{\Omega}_{2}=\frac{1}{64}$ as qualitatively visualized in Fig. \ref{figure:examples:broad:narrow}. Both sources exhibit $\bar{\omega}_{d}=\frac{1}{2}$. Such a configuration can be found, for example, in radio astronomical imaging, where the power of weak broad-band electromagnetic emission from a celestial object is measured while terrestrial radio interference occurs in the observed frequency band \cite{Fridman01}. Fig. \ref{broad:narrow:loss:vs:snr:diff} visualizes the hard-limiting loss where the SNR $\bar{\theta}_{2}$ varies while the SNR $\bar{\theta}_{1}$ is fixed. Results show that the information loss $\chi_1$ increases when changing from $\bar{\theta}_{1}=-15$ dB to $\bar{\theta}_{1}=-3$ dB. In contrast, $\chi_2$ slightly decreases for low SNR $\bar{\theta}_{2}$ while obtaining its minimum around an SNR of $\bar{\theta}_{2}=12$ dB. As for the first scenario with two narrow signal sources, results indicate that SNR imbalance is unfavorable for the weak broad-band source, while the hard-limiting loss caused by high SNR interference with narrow bandwidth is less pronounced.
Fig. \ref{broad:narrow:nrmse:vs:snr} depicts the empirical results for Monte-Carlo simulations of the Fisher scoring method ($I=5$, $\hat{\theta}^{(0)}_d=\theta_\Delta=-30$ dB, $K=10^3$), which are in accordance with the analytical performance analysis.
\section{Conclusion}
We have discussed spectral analysis with binary measurements of random signals under the assumption that the power spectral density structure of the analog sources is known. Using an auxiliary model characterizing multivariate binary distributions, we have obtained the achievable performance and a tractable method for power parameter estimation from the binary samples. The results show that the information loss due to hard-limiting depends on the particular scenario and support that spectral analysis from binary sensor data is, in general, feasible for low to medium SNR configurations when using likelihood-oriented techniques. Future research includes analyzing the potential benefits of oversampling onto the achievable sensitivity of likelihood-oriented spectral parameter estimation, i.e., studying cases where $\Omega_{\text{A/D}}>\Omega_{0}$.

\bibliographystyle{IEEEbib}
\bibliography{IEEEabrv,Bibliography}
\end{document}